\documentclass[conference]{IEEEtran}
\IEEEoverridecommandlockouts
\usepackage{cite}
\usepackage{amsmath,amssymb,amsfonts}
\usepackage{algorithmic}
\usepackage{graphicx}
\usepackage{textcomp}
\usepackage{xcolor}
\usepackage{booktabs}
\usepackage{framed}
\usepackage[flushleft]{threeparttable}
\def\BibTeX{{\rm B\kern-.05em{\sc i\kern-.025em b}\kern-.08em
    T\kern-.1667em\lower.7ex\hbox{E}\kern-.125emX}}
\usepackage{listings}% http://ctan.org/pkg/listings
\begin{document}

\IEEEpubid{\makebox[\textwidth][l]{\begin{minipage}[t]{\columnwidth}\vspace{2pt}\footnotesize
\copyright~2026 IEEE. Personal use of this material is permitted. Permission from
IEEE must be obtained for all other uses, in any current or future media, including
reprinting/republishing this material for advertising or promotional purposes,
creating new collective works, for resale or redistribution to servers or lists,
or reuse of any copyrighted component of this work in other works.
\end{minipage}}}

\title{Robust and Efficient Feature Extraction for Spike Sorting via the Walsh-Hadamard Transform}

\author{\IEEEauthorblockN{Emily L Yang\textsuperscript{1}, Liyuan Guo\textsuperscript{1,2}, Seyed Mohammad Ali Zeinolabedin\textsuperscript{2}, Meng Zhang\textsuperscript{3}, Ke Yang\textsuperscript{3}, \\ Matthieu Couriol\textsuperscript{1,2}, Christian Mayr\textsuperscript{3}, and Pierre-Emmanuel Gaillardon\textsuperscript{1}}
\IEEEauthorblockA{\textit{\textsuperscript{1}University of Utah, Salt Lake City, UT, USA} \\
\textit{\textsuperscript{2}BlackRock Neurotech, Salt Lake City, UT, USA} \\
\textit{\textsuperscript{3}Dresden University of Technology, Dresden, Germany} \\
emily.lyn.yang@utah.edu}
}

\maketitle

\begin{abstract}
Implantable neural interfaces require low-power real-time signal processing to remain within strict thermal and bandwidth constraints, motivating lightweight feature extraction methods for on-chip spike sorting. This work presents the Walsh-Hadamard Transform (WHT) as a hardware-efficient feature extraction method for neural spike classification. WHT can be implemented using only adders, subtractors, and registers without coefficient memory. WHT performance is compared against the Compressed Hadamard Transform (CHT) and Principal Component Analysis (PCA), improving mean F1-scores from 55--60\% to 70--75\% on difficult high-noise datasets and from 90--95\% to 95--99\% on all other simulated datasets. In addition to improved classification performance, WHT demonstrates greater robustness to noise, downsampling, reduced training size, and distance metric selection, maintaining standard deviations typically below 5\%, while CHT and PCA reach up to 10\% under high-noise conditions.
\end{abstract}


\section{Introduction}
Advances in microtechnology have allowed multi-electrode arrays to be manufactured in a scalable manner. These implantable electrode arrays can capture multi-channel recordings from thousands of neurons simultaneously to be used for neuro-prosthetic and research applications. However, increase in electrode count corresponds to higher throughput, memory, and power requirements. The power budget of conventional 1000-channel recording systems of raw neural signals can far exceed the power levels acceptable for implantable devices, where thermal dissipation must be limited to prevent damage to surrounding tissue \cite{zeinolabedin22,kim07,guo24}. Both bandwidth and power considerations for data transmission naturally lead to designing on-chip data processing systems to reduce the data rate and power consumption. 

The data input of many neural applications, such as brain-machine interfaces (BMI), involve signals from individual neurons (i.e. spikes). Spike sorting is the process of separating multi-unit activity from multiple neurons into individual units, and typically consists of spike detection, feature extraction, and clustering/classification stages \cite{gibson12}. Detection is commonly performed using thresholding techniques on preprocessed signals such as NEO/TEO \cite{lewicki98}, while feature extraction reduces dimensionality with methods such as principal component analysis (PCA) and time-frequency transforms \cite{quiroga04}, or lightweight alternatives including geometric \cite{gibson12} or derivative-based features \cite{paraskevopoulou13}. Clustering/classification then groups spikes in feature space to classify each action potential/spike to different neurons.

On-chip incorporation of these processing stages requires them to be low-power, low-area, automatic, and able to operate in real-time for BMI applications. Existing methods to extract spectral band features to discriminate between spikes, such as discrete wavelet transform (DWT), discrete cosine transform (DCT), fast Fourier transform (FFT), or applying finite impulse response (FIR) filters \cite{quiroga04} require many multiply-and-accumulate operations and on-chip storage for coefficients, making them unsuitable for on-chip processing of multi-channel recordings.

This work proposes the Walsh-Hadamard Transform (WHT) as a hardware-efficient feature extraction method for spike sorting purposes. WHT has already been explored in neural signal processing in the context of data compression for signal reconstruction \cite{hosseini-nejad14}, but there exists minimal discussion pertaining to its use for feature extraction in the context of spike sorting. The hardware efficiency of the Compressed Hadamard Transform (CHT) has been previously discussed \cite{uran22}, and this work demonstrates that minimal hardware modifications are needed to implement the WHT. While incurring similar hardware cost to CHT, the WHT provides significant upsides both in terms of algorithmic stability and spike sorting performance. Mean F1-scores improve from 55--60\% to 70--75\% on difficult high-noise datasets and from 90--95\% to 95--99\% on all other simulated datasets. Additionally, WHT maintains standard deviations below 5\% compared to standard deviations up to 10\% for CHT at various noise levels.

In the rest of this paper, Section II introduces WHT along with a proposed hardware implementation. Section III evaluates the performance of WHT against existing feature extraction methods, and Section IV concludes the work.
\begin{figure}[t]
\centerline{\includegraphics[scale=0.15]{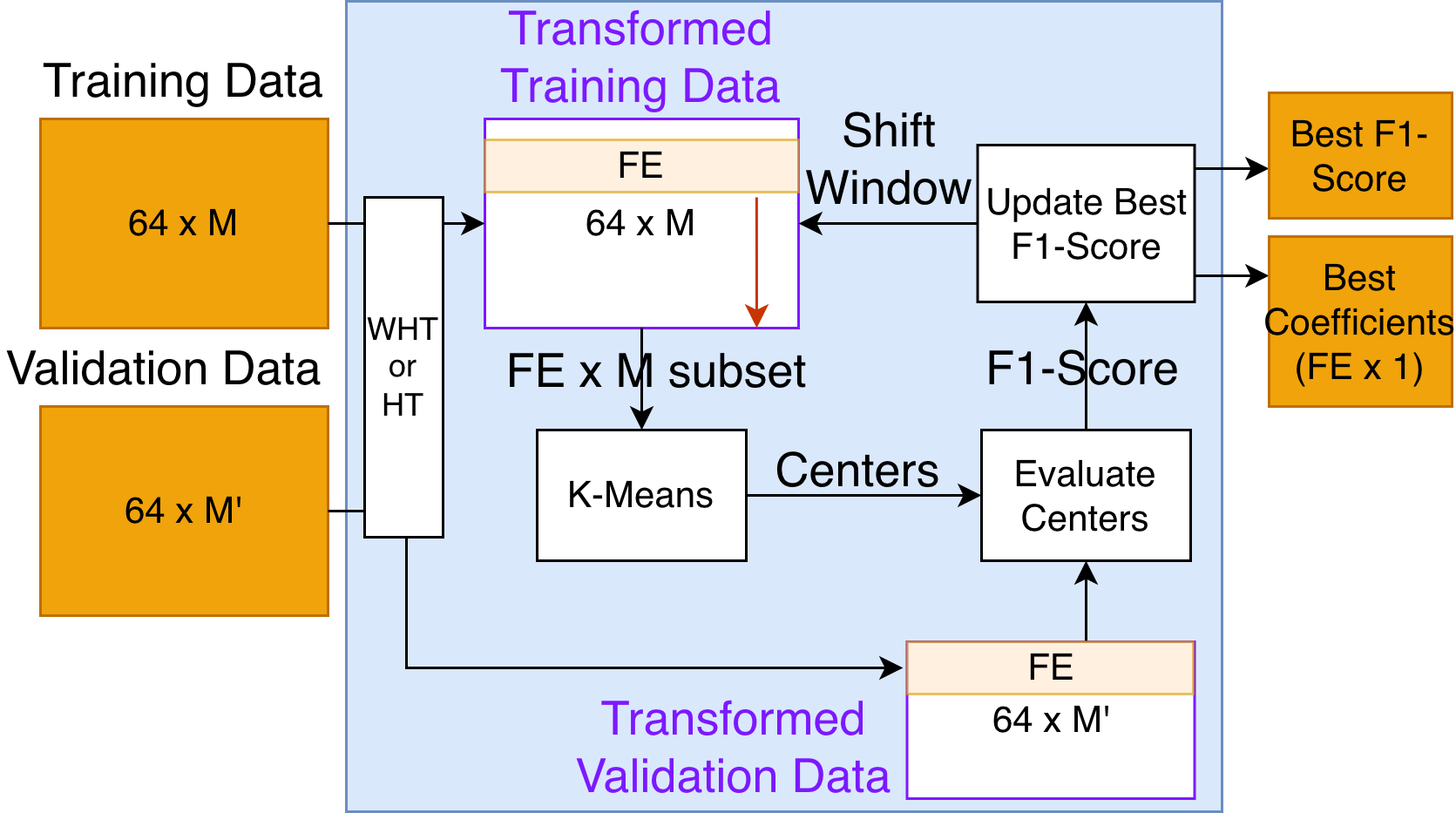}}
\caption{Training workflow for WHT with sliding window coefficient selection scheme.}
\label{train_flow}
\end{figure}

\begin{figure*}[ht!]
\centerline{\includegraphics[scale=0.15]{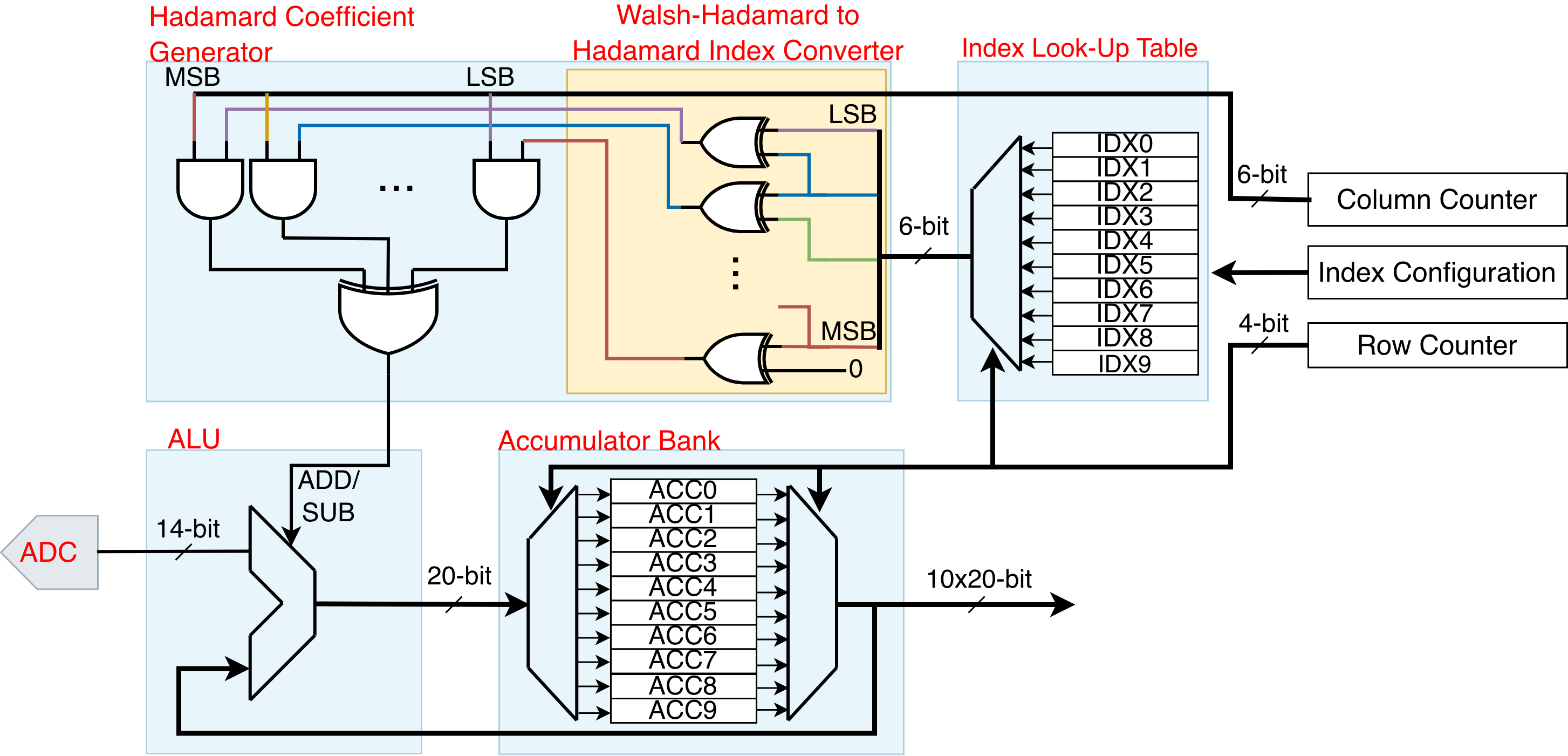}}
\caption{Block diagram of the configurable digital WHT hardware with support for up to 10 coefficients, assuming 14-bit ADC and 64-sample spikes. Yellow block shows the main modification from CHT block diagram in \cite{uran22} to convert from WHT to HT row ordering.}
\label{wht_hw}
\end{figure*}

\section{Proposed Methodology}
\subsection{Walsh-Hadamard Transform}
The WHT is a row-permuted version of the standard Hadamard Transform (HT). The HT matrix follows the recursive Sylvester ordering and can be generated as:
\begin{equation}
    H_0 = [1],\
    H_1 = 
    \begin{bmatrix}
        1 & 1 \\
        1 & -1
    \end{bmatrix},
    H_n =
    \begin{bmatrix}
        H_{n-1} & H_{n-1} \\
        H_{n-1} & -H_{n-1}
    \end{bmatrix}
\label{ht}
\end{equation}

%The WHT preserves the same basis functions as the HT, but reorders the rows according to an increasing number of sign changes, known as `sequency ordering'. 
The WHT matrix \(W_n\) preserves the same basis functions as the HT matrix, but reorders the rows according to an increasing number of sign changes, known as `sequency ordering'. Its relevance to spike sorting is discussed later in this section.
For a given \(H_n\), the corresponding matrix will be \(2^n\times 2^n\). 

%To convert the WHT row index to the HT row index, the binary representation of the WHT row index is first converted to Gray Code and then the resulting bit sequence is reversed. 
To obtain the WHT row index from the HT row index, the binary representation of the HT row index is first bit-reversed, and the resulting bit sequence is then decoded as a Gray-code word.
Since HT coefficients can be generated in-situ, the rows do not need to be reordered explicitly. Instead, the desired row is obtained by remapping the requested WHT row index to the HT index before coefficient generation, allowing the WHT to retain nearly the same hardware complexity as the CHT accelerator in \cite{uran22}. Since basis coefficients are restricted to \(+1\) and \(-1\), 
%implementation is enabled using only adders, subtractors, and registers without requiring coefficient memory. 
implementation uses only adders, subtractors, and registers without coefficient memory.

The input spikes in this work contain 64 samples, thus \(H_6\) and \(W_6\) are used. Mathematically, the transform operation is represented as:

\begin{equation}
    F_{WHT,(64\times 1)} = W_{6(64\times 64)}\times S_{(64\times 1)}
\label{transform_eq}
\end{equation}

\noindent where \(S\) is the input spike waveform and \(F_{\mathrm{WHT}}\) is the transformed feature vector. Each transformed coefficient corresponds to the dot product between the input spike and one Walsh basis row. If intermediate sums are stored, each coefficient can also be computed incrementally as samples stream through the datapath. 

Although the HT and WHT are not true frequency-domain transforms, sequency ordering provides a frequency-like interpretation in which low-sequency coefficients capture slowly varying waveform components, while high-sequency coefficients capture rapidly varying components \cite{beauchamp79}. This ordering is particularly useful for feature extraction because consecutive WHT coefficients correspond to basis functions with similar sequency content. Consequently, selecting consecutive WHT coefficients can be interpreted as selecting a compact sequency band, analogous to selecting a spectral band in conventional frequency-domain methods.

This is relevant for spike sorting, where discriminative information may be localized to specific waveform bands. In contrast, the original HT ordering lacks a monotonic index–sequency relationship. As a result, neighboring HT coefficients do not necessarily correspond to neighboring sequency components, making coefficient subset selection less structured and potentially requiring exhaustive combinatorial search or heuristic ranking \cite{quiroga04}.

Motivated by the structured sequency ordering of the WHT, we propose a sliding-window coefficient selection method, illustrated in Fig. \ref{train_flow}, which evaluates only consecutive sequency bands. This reduces the search complexity from combinatorial in the number of selected features to linear in the spike length \(N\), while preserving an intuitive band-like interpretation of the selected features.

Given \(M\) training spikes of length \(64\), the training dataset has dimensions \(64 \times M\). After applying the WHT to each spike as shown in (\ref{transform_eq}), a window containing the desired number of feature elements is swept across the transformed vectors to generate candidate feature subsets of size \textit{FE}. The first candidate subset contains the first \textit{FE} coefficients from each transformed spike, while the final distinct subset contains the 64th coefficient along with the first $\textit{FE}-1$ coefficients, forming a circular window.

For each candidate subset, k-means clustering is performed on the training data to obtain cluster centers, and the resulting F1-score is evaluated using labeled validation data. The window is then shifted by one coefficient and the process is repeated until all consecutive coefficient combinations have been evaluated. The coefficient subset achieving the highest F1-score is retained.

\subsection{Hardware Implementation}
Fig. \ref{wht_hw} contains the hardware implementation of the sequential WHT computation described in Section II-A for up to FE=10 simultaneously generated coefficients. Each accumulator/output coefficient corresponds to the accumulated dot product between the streamed spike samples and one selected Walsh basis row, which are selected from using a 4-bit row counter control signal. Assuming the AFE outputs 14-bit values and each spike contains 64 samples, the accumulator output width is 20 bits to avoid overflow. Since the Walsh basis coefficients are restricted to +1 and -1, the computational unit performs addition or subtraction depending on the coefficient generator output.

The row indices of the selected coefficients are configured based on the training results described in Section II-A  into the index look-up table using 6-bit addresses to represent any of the 64 Walsh basis rows. The column counter represents the sample index of the incoming spike waveform and iterates from 0 to 63 during each spike acquisition. For every clock cycle, the coefficient generator produces the corresponding \(+1\) or \(-1\) basis value associated with the current sample position and configured row index using bit operations.

In the CHT accelerator from \cite{uran22}, the index look-up table connects directly to the Hadamard Coefficient Generator. The primary modification required to convert the original CHT accelerator into a WHT accelerator is the Walsh-Hadamard to Hadamard Index Converter shown in the highlighted block of Fig. \ref{wht_hw}. This block performs the Gray Code conversion followed by bit reversal to map the requested Walsh row index into its corresponding Hadamard row index. The bit reversal is implemented directly through reversed bit connections, allowing the remapped row index bits to be applied with the column counter bits during coefficient generation.

\section{Results}
\subsection{Performance Analysis Methodology}
To assess the WHT's classification performance with the proposed sliding window methodology, its F1-score using 2 to 10 coefficients will be compared against the CHT using the same sliding window coefficient selection scheme described in Section II and PCA using the same number of components. 

Training data is randomly selected to be 1\%, 10\% or 60\% of the total spikes from 20 different labeled datasets with varying noise levels and distinguishability between classes \cite{quiroga04}. Features of training data are extracted and k-means is used to obtain cluster centers \cite{do19}. The remaining data is used as validation data to evaluate the centers generated by each coefficient combination to select the combination with the highest F1-Score as shown in Fig. \ref{train_flow}. Validation data are sorted to the nearest cluster using either the Manhattan or Euclidean distance metrics and then compared to the corresponding center labels to determine F1-score \cite{guo22}. This process is replicated 20 times with randomized training data selection to evaluate the consistency of the methodology. 

\begin{figure}[t]
\centerline{\includegraphics[scale=0.17]{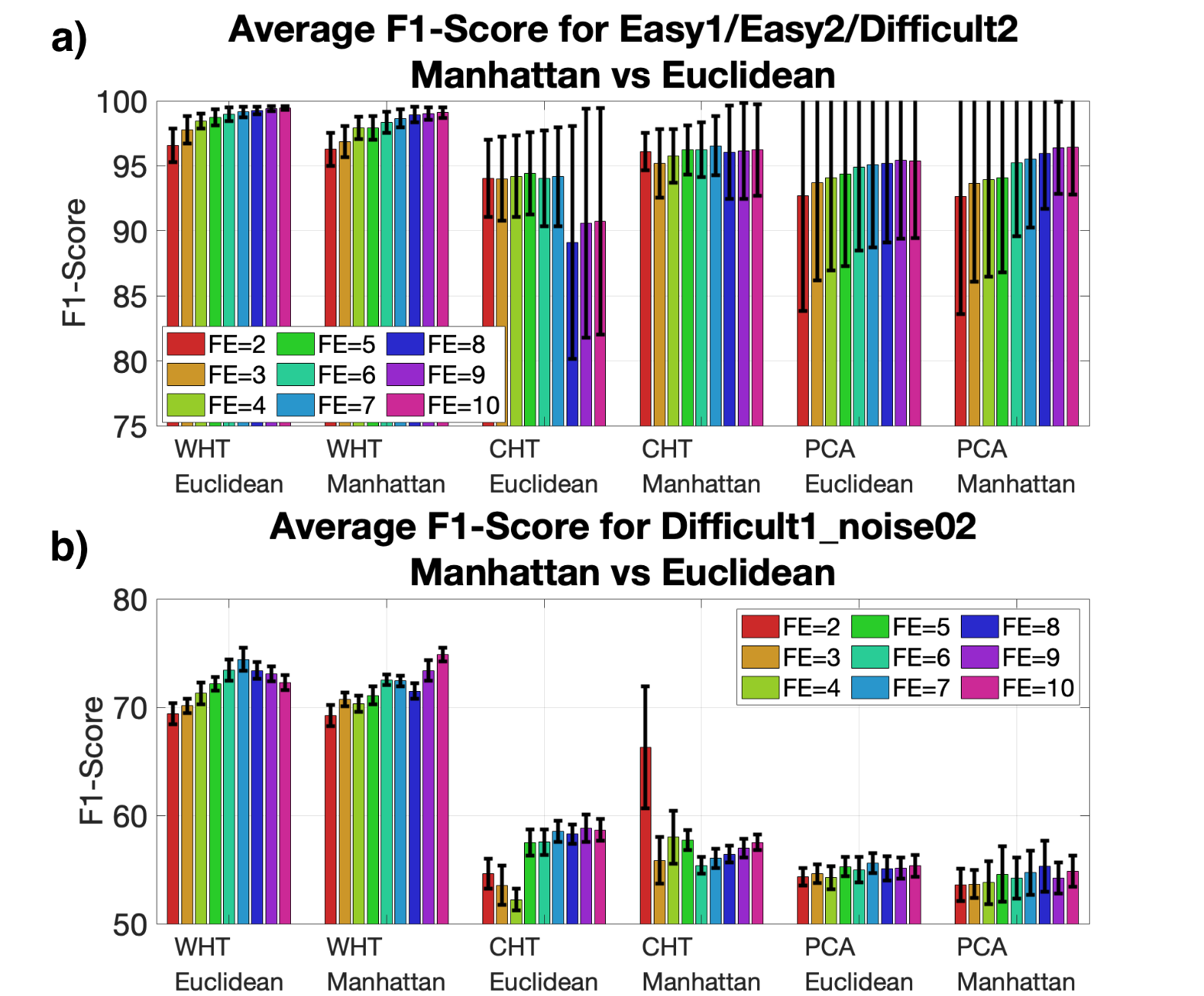}}
\caption{\textbf{a)} Each bar represents the mean F1-score over 320 runs (20 iterations each for 16 datasets) runs using 60\% of data used as training data and the error bars represent one standard deviation. Two different distance metrics used for k-means and validation data assignment are compared. \textbf{b)} Each bar averages 20 runs from the Difficult1\_noise02 dataset.}
\label{dist_methods}
\end{figure}

\begin{figure}[t]
\centerline{\includegraphics[scale=0.17]{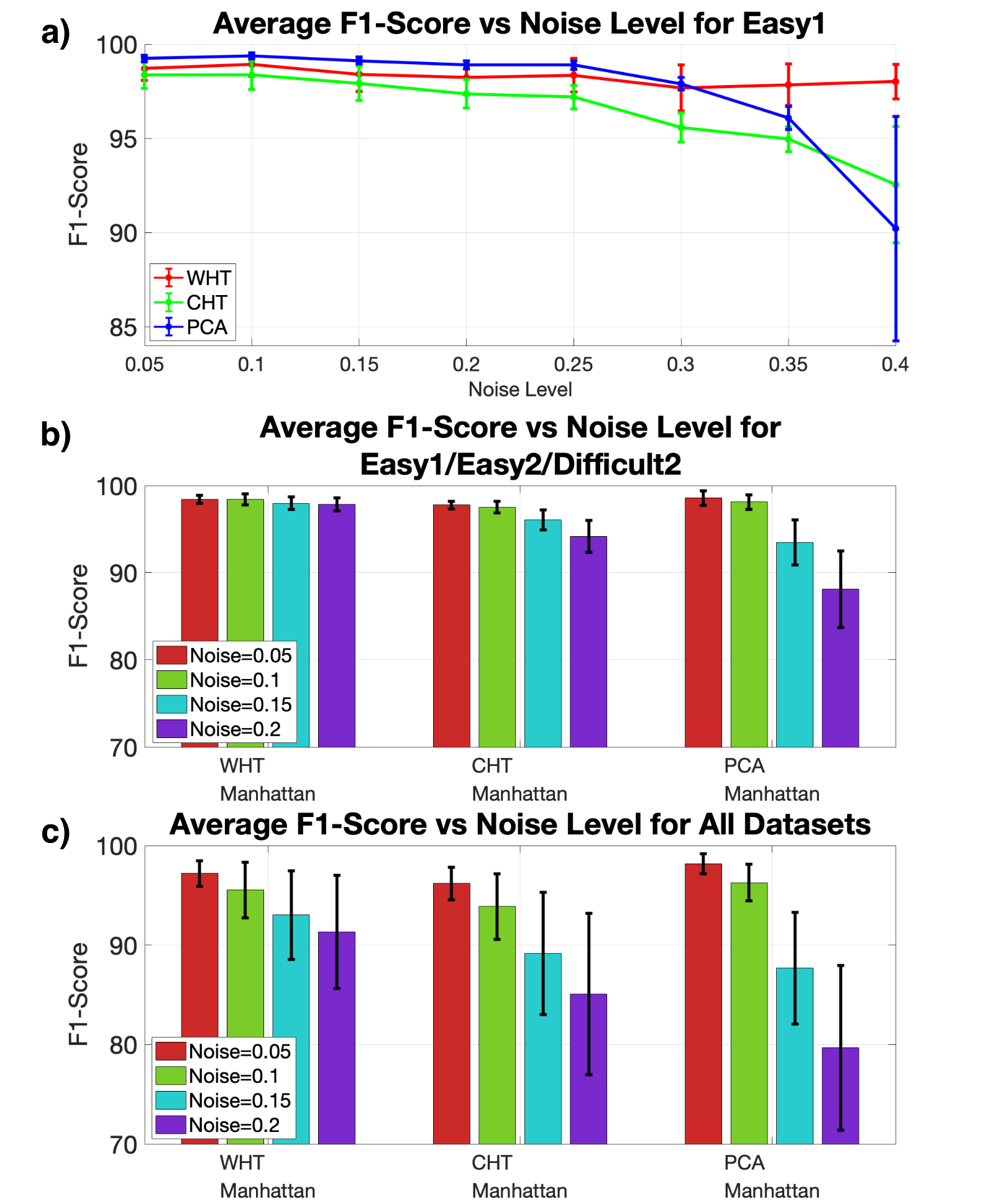}}
\caption{\textbf{a)} Mean F1-score for WHT, CHT, and PCA for all noise levels provided by the Easy1 datasets. \textbf{b)} Mean F1-score from 0.05 to 0.2 for Easy1/Easy2/Difficult2 datasets. \textbf{c)} Mean F1-score from 0.05 to 0.2 noise levels with all datasets included.}
\label{noise_levels}
\end{figure}

\begin{figure}[t]
\centerline{\includegraphics[scale=0.17]{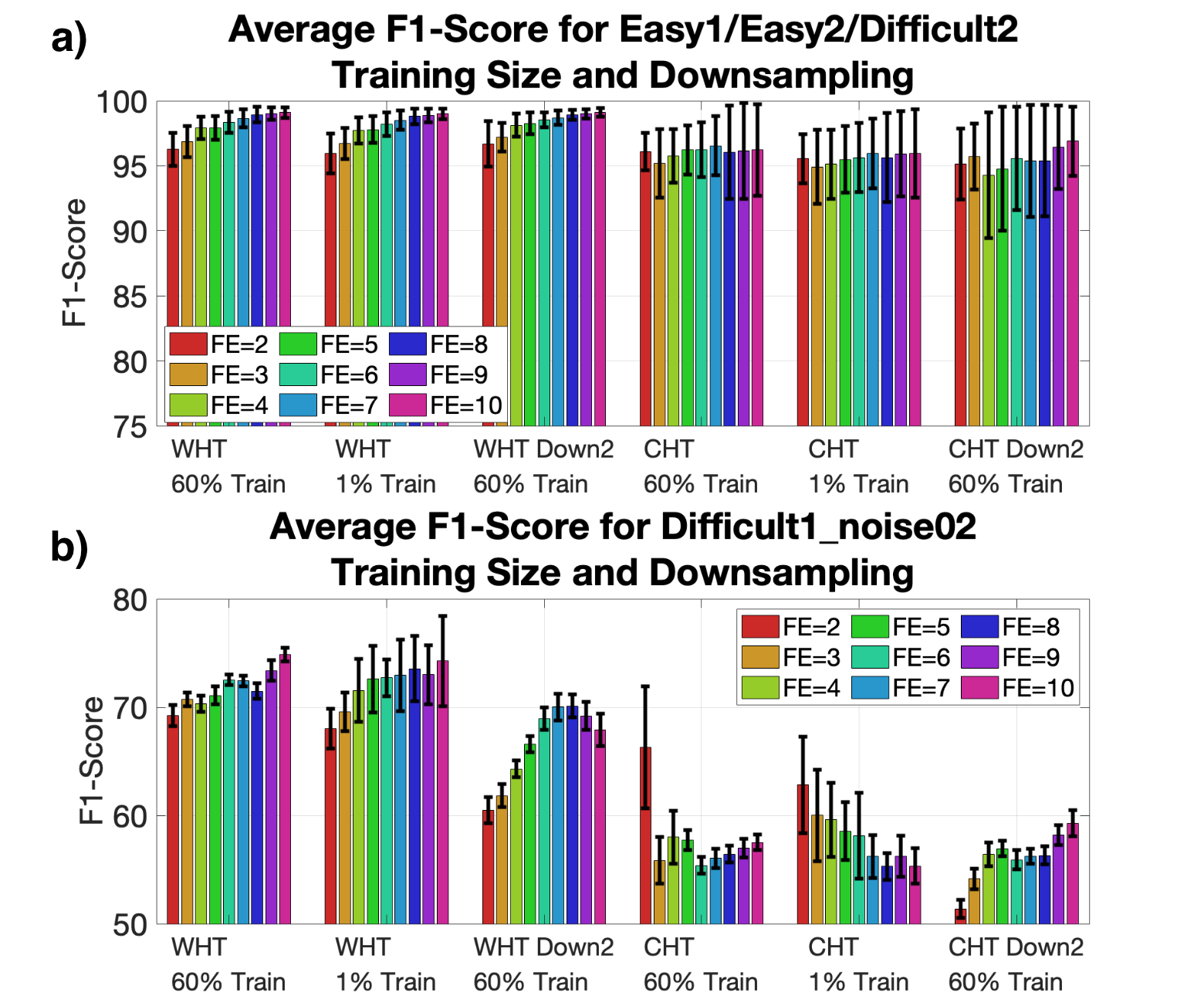}}
\caption{\textbf{a)} Mean F1-score over 320 runs with WHT and CHT using different training set sizes and spike waveform downsampling. \textbf{b)} Each bar averages the F1-score over 20 runs with the same configurations as the top panel on Difficult1\_noise02 dataset.}
\label{train_down}
\end{figure}

\subsection{Comparison against CHT and PCA}
Fig. \ref{dist_methods} compares WHT, CHT, and PCA across varying feature dimensions and distance metrics, while Fig. \ref{noise_levels} evaluates robustness across different noise conditions. Because Difficult1 has substantially lower F1-scores, means are reported separately for Difficult1 and the remaining datasets.

Across all feature dimensions and distance metrics, WHT consistently achieves higher mean F1-scores with lower variance than CHT and PCA. In Fig. \ref{dist_methods}a, PCA and CHT achieve mean F1-scores between 90--95\% with standard deviations ranging from 3--10\%, while WHT maintains 95--99\% mean F1-scores with standard deviations near 1\%. The largest improvement occurs on the Difficult1 dataset in Fig. \ref{dist_methods}b, where WHT improves mean F1-scores from roughly 55--60\% to 70--75\%. Unlike CHT, WHT also exhibits monotonic performance improvements as additional feature elements are included.

Noise robustness results are shown in Fig. \ref{noise_levels}. Although PCA marginally achieves the highest F1-score at low noise levels, both PCA and CHT degrade substantially as noise increases, which leads to lower mean F1-scores and larger standard deviations in Fig. \ref{dist_methods} which averages across noise levels for each dataset. Across Easy1/Easy2/Difficult2 datasets, PCA declines from 98\% to 88\% and CHT from 97\% to 93\% as noise increases to 0.2, while WHT remains within 96--97\%. When all datasets are included, WHT maintains mean F1-scores above 91\% with approximately 3\% standard deviation, whereas CHT and PCA fall to 80--85\% with standard deviations near 5\%.

These results suggest that WHT is less sensitive to noise, feature dimension, and distance metric selection than CHT and PCA, reducing the need for hardware reconfiguration across operating conditions.

\subsection{The Effect of Training Set Size and Downsampling}
Computational load and complexity can be further reduced by decreasing training set size or downsampling spike waveforms. Downsampling from 64 to 32 samples halves both the number transform operations and the number of candidate coefficient windows, enabling support for additional real-time recording channels.

Fig. \ref{train_down} evaluates WHT and CHT using only 1\% of the total data for training and downsampled to 32 samples by using every other sample. In Fig. \ref{train_down}a, WHT maintains nearly identical performance under both modifications, achieving 96--99\% mean F1-scores with low variance. CHT exhibits similar F1-scores, but downsampling increases its standard deviation from roughly 2\% to 4\%. For the Difficult1 dataset in Fig. \ref{train_down}b, both methods show increased variance with smaller training sets; however, WHT still achieves approximately 70\% F1-score for FE=8 and FE=9 with downsampled waveforms, outperforming most CHT configurations at 55--60\%.

\section{Conclusion}
This work presents the WHT with a sliding window to select consecutive coefficients as a robust and hardware-efficient feature extraction method to achieve high spike sorting accuracy. WHT retains the hardware advantages of CHT over FFT, DWT, and FIR-based approaches by eliminating multipliers and coefficient memory, while imposing a more interpretable ordering on the transform coefficients. With the sliding window selection scheme, WHT improves mean F1-scores from 55--60\% to 70--75\% on difficult high-noise datasets and from 90--95\% to 95--99\% on all other simulated datasets. The lower variance across experimental conditions suggests that WHT is less sensitive to hyperparameter selection and noise conditions. 

\newpage
\bibliographystyle{IEEEtran}
\bibliography{refs}

@article{beauchamp79,
   author = {Beauchamp, K. G. and Debnath, L.},
   title = {Walsh Functions and Their Applications},
   journal = {IEEE Transactions on Systems, Man, and Cybernetics},
   volume = {9},
   number = {1},
   pages = {67-67},
   ISSN = {2168-2909},
   DOI = {10.1109/TSMC.1979.4310078},
   year = {1979},
   type = {Journal Article}
}

@article{do19,
   author = {Do, A. T. and Zeinolabedin, S. M. A. and Jeon, D. and Sylvester, D. and Kim, T. T. H.},
   title = {An Area-Efficient 128-Channel Spike Sorting Processor for Real-Time Neural Recording With $0.175~\mu$ W/Channel in 65-nm CMOS},
   journal = {IEEE Transactions on Very Large Scale Integration (VLSI) Systems},
   volume = {27},
   number = {1},
   pages = {126-137},
   ISSN = {1557-9999},
   DOI = {10.1109/TVLSI.2018.2875934},
   year = {2019},
   type = {Journal Article}
}

@article{gibson12,
   author = {Gibson, S. and Judy, J. W. and Marković, D.},
   title = {Spike Sorting: The First Step in Decoding the Brain: The first step in decoding the brain},
   journal = {IEEE Signal Processing Magazine},
   volume = {29},
   number = {1},
   pages = {124-143},
   ISSN = {1558-0792},
   DOI = {10.1109/MSP.2011.941880},
   year = {2012},
   type = {Journal Article}
}

@article{guo24,
   author = {Guo, Liyuan and Weiße, Annika and Zeinolabedin, Seyed Mohammad Ali and Schüffny, Franz Marcus and Stolba, Marco and Ma, Qier and Wang, Zhuo and Scholze, Stefan and Dixius, Andreas and Berthel, Marc and Partzsch, Johannes and Walter, Dennis and Ellguth, Georg and Höppner, Sebastian and George, Richard and Mayr, Christian},
   title = {68-channel neural signal processing system-on-chip with integrated feature extraction, compression, and hardware accelerators for neuroprosthetics in 22 nm FDSOI},
   journal = {Frontiers in Neuroscience},
   volume = {Volume 18 - 2024},
   ISSN = {1662-453X},
   DOI = {10.3389/fnins.2024.1432750},
   year = {2024},
   type = {Journal Article}
}

@inproceedings{guo22,
   author = {Guo, S. and Guo, L. and Zeinolabedin, S. M. A. and Mayr, C.},
   title = {Various Distance Metrics Evaluation on Neural Spike Classification},
   booktitle = {2022 IEEE Biomedical Circuits and Systems Conference (BioCAS)},
   pages = {554-558},
   ISBN = {2163-4025},
   DOI = {10.1109/BioCAS54905.2022.9948670},
   year = {2022},
   type = {Conference Proceedings}
}

@article{hosseini-nejad14,
   author = {Hosseini-Nejad, H. and Jannesari, A. and Sodagar, A. M.},
   title = {Data Compression in Brain-Machine/Computer Interfaces Based on the Walsh–Hadamard Transform},
   journal = {IEEE Transactions on Biomedical Circuits and Systems},
   volume = {8},
   number = {1},
   pages = {129-137},
   ISSN = {1940-9990},
   DOI = {10.1109/TBCAS.2013.2258669},
   year = {2014},
   type = {Journal Article}
}

@article{kim07,
   author = {Kim, S. and Tathireddy, P. and Normann, R. A. and Solzbacher, F.},
   title = {Thermal Impact of an Active 3-D Microelectrode Array Implanted in the Brain},
   journal = {IEEE Transactions on Neural Systems and Rehabilitation Engineering},
   volume = {15},
   number = {4},
   pages = {493-501},
   ISSN = {1558-0210},
   DOI = {10.1109/TNSRE.2007.908429},
   year = {2007},
   type = {Journal Article}
}

@article{lewicki98,
   author = {Lewicki, Michael S.},
   title = {A review of methods for spike sorting: the detection and classification of neural action potentials},
   journal = {Network: Computation in Neural Systems},
   volume = {9},
   number = {4},
   pages = {R53-R78},
   ISSN = {0954-898X},
   DOI = {10.1088/0954-898X_9_4_001},
   year = {1998},
   type = {Journal Article}
}

@article{paraskevopoulou13,
   author = {Paraskevopoulou, Sivylla E. and Barsakcioglu, Deren Y. and Saberi, Mohammed R. and Eftekhar, Amir and Constandinou, Timothy G.},
   title = {Feature extraction using first and second derivative extrema (FSDE) for real-time and hardware-efficient spike sorting},
   journal = {Journal of Neuroscience Methods},
   volume = {215},
   number = {1},
   pages = {29-37},
   ISSN = {0165-0270},
   DOI = {https://doi.org/10.1016/j.jneumeth.2013.01.012},
   year = {2013},
   type = {Journal Article}
}

@article{quiroga04,
   author = {Quiroga, R. Quian and Nadasdy, Z. and Ben-Shaul, Y.},
   title = {Unsupervised Spike Detection and Sorting with Wavelets and Superparamagnetic Clustering},
   journal = {Neural Computation},
   volume = {16},
   number = {8},
   pages = {1661-1687},
   ISSN = {0899-7667},
   DOI = {10.1162/089976604774201631},
   year = {2004},
   type = {Journal Article}
}

@article{uran22,
   author = {Uran, A. and Ture, K. and Aprile, C. and Trouillet, A. and Fallegger, F. and Revol, E. C. M. and Emami, A. and Lacour, S. P. and Dehollain, C. and Leblebici, Y. and Cevher, V.},
   title = {A 16-Channel Neural Recording System-on-Chip With CHT Feature Extraction Processor in 65-nm CMOS},
   journal = {IEEE Journal of Solid-State Circuits},
   volume = {57},
   number = {9},
   pages = {2752-2763},
   ISSN = {1558-173X},
   DOI = {10.1109/JSSC.2022.3161296},
   year = {2022},
   type = {Journal Article}
}

@article{zeinolabedin22,
   author = {Zeinolabedin, S. M. A. and Schüffny, F. M. and George, R. and Kelber, F. and Bauer, H. and Scholze, S. and Hänzsche, S. and Stolba, M. and Dixius, A. and Ellguth, G. and Walter, D. and Höppner, S. and Mayr, C.},
   title = {A 16-Channel Fully Configurable Neural SoC With 1.52 $\mu$W/Ch Signal Acquisition, 2.79 $\mu$W/Ch Real-Time Spike Classifier, and 1.79 TOPS/W Deep Neural Network Accelerator in 22 nm FDSOI},
   journal = {IEEE Transactions on Biomedical Circuits and Systems},
   volume = {16},
   number = {1},
   pages = {94-107},
   ISSN = {1940-9990},
   DOI = {10.1109/TBCAS.2022.3142987},
   year = {2022},
   type = {Journal Article}
}
\end{document}